\documentclass[aps,showpacs,preprint]{revtex4-1}
\usepackage{graphicx}

\newcommand{\beq}{\begin{equation}}
\newcommand{\eeq}{\end{equation}}
\newcommand{\beqa}{\begin{eqnarray}}
\newcommand{\eeqa}{\end{eqnarray}}
\newcommand{\eps}{\epsilon}
\newcommand{\rr}{{\bf r}}
\newcommand{\dd}{{\bf \delta}}
\newcommand{\s}{\sigma }
\newcommand{\p}{{\bf p}}
\newcommand{\Q}{{\bf Q}}
\newcommand{\V}{{\bf V}}
\begin{document}

\title{Stripes and superconductivity in
the two-dimensional self-consistent model }
\author{S. I. Matveenko }
\affiliation{
Landau Institute for Theoretical Physics, Kosygina Str. 2,
119334, Moscow, Russia}
\author{S. I. Mukhin}
\affiliation{Theoretical Physics Department, Moscow Institute for Steel and Alloys, 119991 Moscow, Russia. }
\author{F. V. Kusmartsev}
\affiliation{Department of Physics, Loughborough University, Loughborough, LE11 3TU, UK}

\date{\today}

\begin{abstract}
We found  solutions of the Bogoliubov-de Gennes equations
for the two-dimensional self-consistent model of superconductors with
$d_{x^2-y^2}$ symmetry of the order parameter,
taking into account spin and charge distributions.
Analytical solutions for spin-charge density wave phases
in the absence of the superconductivity ("stripe" and "checkerboard"
structures) are presented. Analytical solutions for  coexisting
superconductivity and stripes  are  found.
\end{abstract}

\maketitle

\section{Introduction}
The accumulating experimental evidence suggests that the pseudogap phase could be a key issue in understanding the underlying mechanism of high-transition-temperature
superconducting (high-T$_c$) copper oxides \cite{taill}. Magnetotransport data in electron-doped copper oxide La$_{2-x}$Ce$_x$CuO$_4$ suggests that linear temperature-dependent
resistivity correlates with the electron pairing and spin-fluctuating scattering of the electrons \cite{greene}. Simultaneously, in the hole-doped copper oxide YBa$_2$Cu$_3$O$_y$ a large in-plane
anisotropy of the Nernst effect sets in at the boundary of the pseudogap phase and indicates that this phase breaks four-fold rotational symmetry in the a-b plane, pointing to stripe or nematic order \cite{taill1}.
Hence, we study here analytically a coexistence of the stripe order and d-wave superconductivity. We found an exact solution of the Bogoliubov-de Gennes equations in the simple Hubbard
t-U-V mode indicating that the Abrikosov's vortex core naturally gives rise to a stripe-ordered domain. We show that the size of the stripe domain may exceed superconducting vortex's core size $\xi_s$ and the inter-vortex distance in the Abrikosov's lattice in the limit of weak magnetic fields $H \leq H_{c1}$. As far as we know, this is the first analytic solution of such type. Previously, coexistence of the stripe-order and Abrikosov's vortices in the limit of high magnetic fields $H\sim H_{c2}$ has been investigated numerically \cite{knapp}. Calculations were limited by the size of the model cluster of 26$\times$52 sites. Hence, the numerical results covered only the case when the inter-vortex distance was less than correlation length of the static AFM order. Here we consider analytically the opposite case of weak magnetic fields, where the inter-vortex distance is mach greater than stripe-order (including AFM) correlation length.
\%
\%
Predicted numerically stripe order \cite{Zaanen}, e.g. coupled spin- and charge-density periodic superstructure (SDW-CDW), was found in the underdoped superconducting
cuprates experimentally, specifically in $La_{2-x}Ba_xCuO_4$ \cite{fujita} and $La_{1.6-x}Nd_{0.4}Sr_xCuO_4$ \cite{ich,hu}.  It was shown analytically, that stripe-order may arise already in the short-range repulsive Hubbard model due to a quantum interference between backward and Umklapp scattering of electrons by SDW potential close to half-filling in the presence of the CDW order with "matching" wave-vector \cite{sm}. In the quasi-1D case analytical kink-like spin- and charge-density coupled solutions  were found \cite{mm} in the normal state. A study of $La_{1.875}Ba_{0.125}CuO_4$ with angle-resolved photoemission and scanning tunneling spectroscopies \cite{valla} has found evidence for a d-wave-type gap at low temperature, well within the stripe-ordered
phase but above the bulk superconducting Tc. An earlier  inelastic neutron scattering data \cite{aeppli1} had shown field-induced fluctuating magnetic order with space periodicity $8a_0$ and wave vector
pointing along Cu-O bond direction in the ab-plane of the optimally doped
La$_{1.84}$Sr$_{0.16}$CuO$_4$ in external magnetic field of $7.5$ T
below $10$ K. The applied magnetic field ($\sim 2-7$ T)
imposes the vortex lattice and induces "checkerboard" local density of
electronic states (LDOS) seen in the STM experiments in high-T$_c$
superconductor Bi$_2$Sr$_2$Ca Cu$_2$O$_{8+\delta}$ \cite{davis}. The pattern
originating in the
Abrikosov's vortex cores has $4a_0$ periodicity, is oriented along  Cu-O
bonds, and has decay length $\sim 30$ angstroms
 reaching well outside the vortex core.
The existence of antiferromagnetic spin fluctuations well outside the vortex
cores is also discovered by NMR \cite{nmr} in superconducting YBCO in a $13$ T
external magnetic field.
Theoretical predictions had also been made of the magnetic field induced
coexistence of antiferromagnetic ordering phenomena and superconductivity in
high-T$_c$
cuprates \cite{zhang,arovas,sachdev1,sachdev2,sachdev3} due to assumed proximity of pure superconducting state to a
phase with co-existing superconductivity and spin density wave order.
 In these works effective Ginzburg-Landau
theories of coupled superconducting-, spin- and charge-order fields were used.
Alternatively, the fermionic quasi-particle weak-coupling approaches were
focused on the theoretical predictions arising from the model of BCS
superconductor with $d_{x^2-y^2}$ symmetry \cite{volovik}. An effect of the
nodal fermions on the zero bias conductance peak in tunneling studies was
predicted. However STM experiments of the vortices in high-T$_c$ compounds
revealed a very different structure of LDOS \cite{davis}.
In this paper we make an effort to combine both theoretical approaches and
present analytical mean-field solutions of coexisting spin-, charge- and
superconducting orders derived form microscopic Hubbard model in the
weak-coupling approximation. The  previous analytical results obtained in
the quasi 1D cases \cite{mm,mm1,ma,fe} are now extended for two real
space dimensions. Different analytical solutions for collinear and checkerboard
stripe-phases, as well as for spin-charge density modulation inside
Abrikosov's vortex core are obtained. Simultaneously, our theory provides
wave-functions of the fermionic states in all considered cases.

\section{ Effective hamitonian. Bogoliubov-de Gennes equations}

Consider   the Hamiltonian $H = H_0 + H_{sc}$ consisting of two parts:
the first part is the   Hubbard Hamiltonian with  on-site repulsion
$U > 0$

\begin{equation}
H_0 =\displaystyle -t\sum_{\langle i,j\rangle,
\sigma}c^{\dagger}_{i,\sigma}c_{j,\sigma}+ U\displaystyle\sum_{i}
\hat{n}_{i,\uparrow} \hat{n}_{i, \downarrow}
- \mu \sum_{i. \sigma} \hat{n}_{i, \sigma} ,
  \label{hubbard}
\end{equation}
and the interaction part including superconducting correlations
\beq
H_{sc} = \sum_{<i,j>, \s} \Delta(i,j;\sigma)c^{\dagger}_{i,\sigma}
c^{\dagger}_{j,-\sigma} + h.c.,
\label{hs}
\eeq
where $\sum_{<i,j>, \s}$ is a summation over nearest neighboring sites
${\bf r}_i$, ${\bf r}_j$ of the square lattice, and spin  components $\s = 2 s_z = \pm 1$.

In the self-consistent approximation the Hamiltonian acquires the form
\beqa
H = -t \sum_{\langle i,j\rangle
\sigma}c^{\dagger}_{i,\sigma}c_{j,\sigma}  +\frac{U}{2}
\sum_{i,\sigma}(\rho_i c^{\dagger}_{i,\sigma}c_{i,\sigma} -\frac{\rho_i^2}{2})
\nonumber \\
-U \sum_{i, \sigma} \langle \hat{S}_z (\rr_i ) \rangle \sigma
 c^{\dagger}_{i,\sigma}c_{i,\sigma} +U \langle \hat{S}_z (\rr_i ) \rangle^2
-\mu \sum_{i, \sigma}c^{\dagger}_{i,\sigma}c_{i,\sigma} \nonumber \\
\sum_{<i,j>,\s} \Delta(i,j;\sigma)c^{\dagger}_{i,\sigma}
c^{\dagger}_{j,-\sigma} +h.c. + \frac{|\Delta |^2}{g},
\label{H}
\eeqa
where we introduce similar \cite{mm} slowly varying functions for spin order parameter
$m(\rr_i )$
and the charge density $\rho (\rr_i)$ defined as
\beqa
\rho (\rr) = \langle \hat{n} (\rr ) \rangle, \quad
(-1)^{x_i + y_i} m (\rr_i ) =
 U \langle \hat{S}_z (\rr_i ) \rangle, \nonumber \\
\Delta(i,j;\sigma)= -g \langle c_{j,-\sigma}
c_{i,\sigma}\rangle.
\label{selfc}
\eeqa
We can diagonalize the total Hamiltonian $H = H_0 + H_{sc}$ by performing a
 unitary Bogoliubov transformation
\beq
\hat{c}_{\sigma}(\rr) = \sum_n \gamma_{n, \sigma} u_{n, \sigma}(\rr)-\sigma
\gamma^+_{n, -\sigma} v^*_{n, -\sigma}(\rr)
\label{tr}
\eeq
New operators $\gamma$, $\gamma^+$ satisfy the fermionic commutative
relations $
\{\gamma_{n, \sigma}, \gamma^+_{m, \sigma^{\prime}}\} = \delta_{m,n}
\delta_{\sigma, \sigma^{\prime}}$.
The transformations (\ref{tr}) must diagonalize the Hamiltonian $H$:
\beq
H =  E_g + \sum_{\eps_n >0} \eps_n \gamma_{n,\sigma}^+ \gamma_{n,\sigma},
\label{eg}
\eeq

where $E_g$ is the ground state energy and $\eps_n >0$ is the energy
of  the n-th excitation.
Following \cite{bdg} we obtain the eigenvalue equations
\beqa
-t\sum_{\dd} u_{\s}(\rr +\dd )+ (\frac{U}{2} \rho(\rr)-\mu )u_{\s}&
+m (\rr )(-1)^{x_i +y_i} \s u_{\s}(\rr ) \nonumber \\
+\sum_{\dd} \Delta(\rr,\rr +\dd;\s )\s v_{\s} (\rr +\dd )&
 =\eps_{\s} u_{\s}(\rr),
 \label{deq1}
\eeqa
\beqa
-\sum_{\dd} \Delta^*(\rr,\rr +\dd;-\s )\s u_{\s} (\rr +\dd )
+t\sum_{\dd} v_{\s}(\rr +\dd ) \nonumber \\
- (\frac{U}{2} \rho(\rr)-\mu )v_{\s}
+m (\rr )(-1)^{x_i +y_i} \s v_{\s}(\rr ) = \eps_{\s} v_{\s}(\rr),
\label{deq2}
\eeqa
where $\dd=\pm \hat{\bf x}, \pm \hat{\bf y}$.

We suppose the $d_{x^2 -y^2}$ symmetry of the
 superconducting order parameter $\Delta(\rr ,\rr \pm \hat{\bf x};\s)=
 \s \Delta_d (\rr )$, $\Delta(\rr ,\rr \pm \hat{\bf y};\s)=
 -\s \Delta_d (\rr )$. The Fourier transform gives the usual dependence
 $\Delta_{\p} (\rr) =$
$
 \s \sum_{\dd} \Delta_{sc}(\rr,\rr+\dd;\s)
\exp[-i \p \dd ]= 2(\cos p_x - \cos p_y ) \Delta_d (\rr).
$
The system (\ref{deq1}) - (\ref{deq2}) can be rewritten in the continuum approximation.
Consider states near the Fermi surface (FS)  (see Fig.1)
 and use linear
approximation for the quasiparticles spectrum.
Since for SDW pairing components with wave vectors
${\bf p}$ and ${\bf p} - {\bf Q_+}$ (or
 ${\bf p}$ and ${\bf p} - {\bf Q_-}$, where
 ${\bf Q_+} -{\bf Q_-} = 2\pi (0,1)$ is the lattice vector for the
 pure system without doping, when $(-1)^{x_i + y_i} \equiv e^{\pm i \Q \rr}$. )
 are important (see Fig. 1),
 we represent the functions $u(\rr )$ and $v(\rr )$, similar to the
 one-dimensional case, as
\beq
u_{\s}(\rr ) =\sum_{\p \in FS, p_x >0}[u_{\p, \s} e^{i \p \rr}
+ \s u_{\p - \Q, \s} (\rr ) e^{i(\p -\Q )\rr }],
\eeq
where $\Q = \Q_+$ for wave vectors $\p_y > 0$ and $\Q = \Q_-$ for wave vectors
$p_y < 0$, respectively.

 \begin{figure}[tbph]
\begin{center}
\includegraphics[width=3.0in]{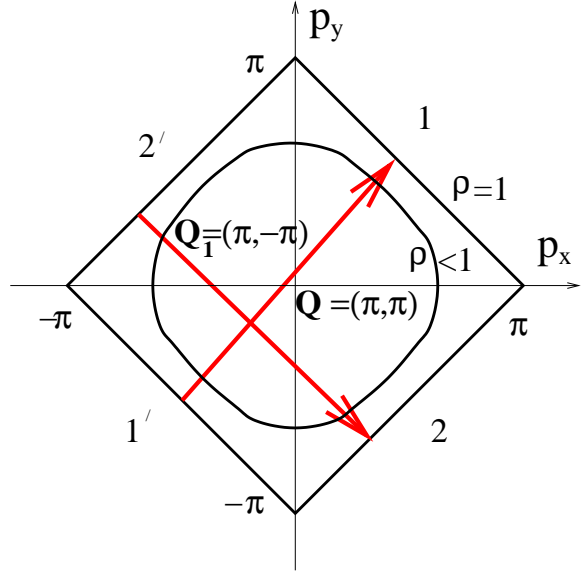}
\caption{ The Fermi surface }
\end{center}
\end{figure}

For the doped case nesting vectors $ \Q_{\pm}$ are no longer equivalent.
Therefore in the general case we consider vectors $\Q_{\pm}$ as independent and
make the substitution $(-1)^{x_i + y_i} m (\rr_i ) \to$
\[
 m_+ (\rr_i )\exp (i \Q_+
\rr_+ ) + m_- (\rr_i )\exp (i \Q_- \rr_- ) + h.c.
\]

Eigenvalue equations (\ref{deq1}), (\ref{deq2})
 take form similar to the 1D case: $\hat{H}\Psi = \eps \Psi$, with

\beq
\hat{H}=\left( \matrix{ A_{\p} & B_{\p} \cr
                                         B_{-\p} & -A_{\p}}\right),
\quad B_{\p} = \left(\matrix{\Delta_{-\p} &0 \cr
                   0 & \Delta_{-\p +\Q}}\right),
\label{HH}
\eeq
\beq
A_{\p} = \left(\matrix{-iV_{\p}\nabla_{\rr} +\eps_{\p} -\eta&
 m_{\pm} (\rr)\cr
m^*_{\pm} (\rr)  & -iV_{\p-\Q}\nabla_{\rr} +\eps_{\p -\Q} -\eta}\right),
\label{ca}
\eeq
where $\eta(\rr) = \mu - \frac{U}{2}\rho(\rr )$,
$\Psi^T = (u_{\p}, u_{\p-\Q}, v_{\p}, v_{\p-\Q}) =(u_+, u_-, v_+, v_- )$,
$\eps_{\p} = -2t(\cos p_x + \cos p_y ) -\mu$,
${\bf V}_{\p}=2 t (\sin p_x, \sin p_y )$, and,
as before, $\Q = \Q_+$ for wave vectors $\p_y > 0$ and $\Q = \Q_-$ for wave vectors
$p_y < 0$.The sign in $m_{\pm}$ is taken  by the same rule.

In the presence of the magnetic field ${\bf H}$ functions $u, v$ becomes spin-dependent,
and equations (\ref{HH}), (\ref{ca})  are changed: $\nabla \to \nabla - i \frac{e}{c} {\bf A}$, $\mu \to
\mu_{\sigma} =\mu + \sigma H$. For the constant magnetic field perpendicular to the plane
we have ${\bf A} = {\bf H} \times{ \bf r} /2$, so that $\partial_x \to \partial_x +i(e/c)yH/2$,
   $\partial_y \to \partial_y -i(e/c)xH/2$.
We suppose  everywhere that the magnetic field is small $H \sim H_{c_1} \ll H_{c_2}$, ($\xi \ll \lambda$), therefore ignore the effect of terms with vector-potential on the solution
at distances $r \ll \lambda$.

For the case $d_{x^2 -y^2}$ symmetry we consider $\Delta_{-\p}=
\Delta_{\p}=-\Delta_{\p -\Q}= 2(\cos p_x - \cos p_y ) \Delta_d
(\rr)$, which corresponds to $\s \Delta(\rr,\rr \pm
\hat{x};\s) =-\s \Delta(\rr,\rr \pm \hat{y};\s) = \Delta_0$
in the uniform ground state. We retained the main terms in the
expansion over $\Delta$. In the higher order approximation,
instead of the terms $\Delta_{-\p}$ and $\Delta_{-\p + \Q}$  we
would have to write  $\Delta_{-\p}
-i(\nabla_{\p}\Delta_{-\p})\nabla_{\rr}$ and $\Delta_{-\p +\Q} -
i(\nabla_{\p} \Delta_{-\p + \Q})\nabla_{\rr}$. The continuum
approximation is not valid for a band filling very close to the
half-filled case (the number of particles per one site $\rho =1$),
where the Fermi velocity tends to zero at points $\p = (0, \pm
\pi), (\pm \pi, 0)$.

In the homogeneous case $\rho(x)=$ const, $m , \Delta_d = const$
for coexisting spin-
and superconducting order parameters, so
the eigenvalue spectrum has the form

\beq
E^2 = (\sqrt{m^2 + \eps^2 (\p) } \pm \eta )^2 + \Delta_{\p}^2,
\eeq
with $\Delta_{\p} = \Delta_d (\cos p_x - \cos p_y )$.

The self-consistent conditions
are derived  by substitution  of functions $u$, $v$ into (\ref{selfc}),
 similar to the one-dimensional case. In the continuum approximation
 they read:
\beq
\rho(\rr) = 2\sum_{\eps} [(u_+^* u_+ + u_-^* u_- )f +
 (v_+^* v_+ + v_-^* v_- )(1-f)]
\label{s1}
\eeq

\beq (-1)^{x_i + y_i} m (\rr) = 4 U[\sum_{\eps} u_-^* u_+ f - \sum_{\eps} v_-^* v_+
(1-f)] \label{s2} \eeq

\beqa
\Delta_{\bf q}(\rr) =2g\sum_{\eps} (v_{+}^* u_{+} -v_{-}^* u_{-})
[(1-f)(\cos (p_x-q_x)  \nonumber \\
+\cos (p_y-q_y) )-
f ((\cos (p_x + q_x) +\cos (p_y + q_y ) ) ],
\eeqa
where $f = {1}/({\exp[\eps /T] +1})$.
 We omitted spin indices since in our
representation for wave functions all equations are diagonal over spin.

\section{Spin-Charge Density Wave Structures}

In the low doping limit  the ground state of the model is
the periodic charge-spin superstructure with the absence of
superconductivity: $\Delta \equiv 0$.
Consider different structures, having close ground state energies.
In real systems the exact ground state must be determined
by taking into account real long-distance 3D interactions.

\subsection{Diagonal and vertical  stripes}

For diagonal stripes we search the solution in the form  :
\[
 u_{\p}(\rr) =  {u}_{\p} (r_+ ),\quad v_{\p}(\rr) =  {v}_{\p} (r_+ ),
 \]
 where $r_{\pm} = (x \pm y)/\sqrt{2}$, $\p \in FS $.
Substituting into Eqn. (\ref{ca}) we obtain a one-dimensional
eigenvalue equation
\beqa
-iV_p \frac{\partial}{\partial r_+} u_+  +\frac{U}{2}\rho (r_+)u_+
 + m(r_+) u_- = E u_+, \nonumber \\
m^* (r_+) u_+ + i V_p \frac{\partial}{\partial r_+} u_-
  +\frac{U}{2}\rho (r_+)u_- = E u_-,
  \eeqa
where $V_p = 2t\sin p_x$.
The only difference from the considered  one-dimensional model\cite{mm}
is the dispersion of the velocity $V_p$.
This system is exactly solvable.
In the ground state, at $\rho = 1$, we have $m (r_+) = m_0$.
Increased doping leads to the stripe structure.
The one stripe solution
has the form
\beq
m (r_+) = m_0 \tanh \frac{r_+}{\xi},
\label{discrete}
\eeq
where the width $\xi$ is defined from the minimum of the
total energy. Solution (\ref{discrete}) corresponds to $\rho=1$ in the
thermodynamic limit for number of  holes per lattice site.
In our case (\ref{discrete}) is valid only in the vicinity of each single stripe
that enters a periodic superstructure called stripe-phase (compare \cite{schulz}).
Distinct from the Peierls
model,  where $\xi = V_F/m_0$, $V_F = const$, the present model has a more
complicated spectrum. Besides continuum bands $E^2 = V_{k}^2 k_{\parallel}^2
+ m_0^2$ we find some discrete levels (for a given $p_x$) inside the gap:
\beq
E_n^2  = m_0^2 \lambda n(2-\lambda n),
\label{En}
\eeq
where $n$ is integer number, $0\leq n \leq 1/\lambda$, and $\lambda
=\lambda (p_x) =V_p /(\xi m_0)$, or
$\lambda = V_p/\bar{V}$ with $\xi \equiv \bar{V}/m_0$.
Each level inside a gap forms a band due to dispersion
of the coefficient $\lambda (p_x)$.
For $\lambda \geq 1$ we obtain only one level,
$E=0$, with  wave function:
\beq
\psi_{\pm} = (u_+ \pm u_-)/\sqrt{2} \propto \frac{1}{(\cosh r_+/\xi )^{1/\lambda }}.
\label{uv}
\eeq
The wave functions of all states
 are described in terms of the hypergeometric
function $F(a,b |c| z )$, and for local levels
they have polynomial form:
\beq
\psi_{\pm, n} \sim\frac{1}{(\cosh x/\xi )^{1/\lambda -n}}
F[\frac{2}{\lambda} +1 -n, \, -n, \,  \frac{1}{\lambda}-n +1,\,
 \frac{1}{2}(1 + \tanh \frac{x}{\xi} )]
\label{uvn}
\eeq
For $1/2 <\lambda < 1$ two levels $n=0,1$, can already  exist.
Allowing for equality $\lambda = V_p/\bar{V}$ we conclude that
both possibilities $\lambda <1$ and $\lambda >1$ take place, each one
in the proper interval of $p_x$.
Similar to the 1D case \cite{mm} the appearance of the
kink in the spin channel is accompanied by
the local charge distribution
$\rho (r_+) - \langle \rho \rangle \sim
1/\cosh^2(r_+ /\xi )$.
An increase of the doping leads to the periodic spin-charge
density superstructure.
In the limiting case  of "overdoping" ($\mid \rho - 1 \mid
\gg 1/\xi$) the spin-charge structure becomes harmonic
\[
m (r_+) \propto \sin (\pi \mid \rho - 1 \mid r_+), \,
\rho(r_+) - \langle \rho \rangle \propto \cos(2 \pi \mid \rho -1\mid r_+).
\]
For vertical stripes we use an ansatz:
\[
 u_p(\rr) = u_p(x ), \quad v_p (\rr ) = v_p (x)
 \]
and  obtain a  system
of equations with $V_p =2t\sin p_x$, which is similar to the diagonal case.
For the same values $m_0$, $\xi$, the parameter
$\lambda$ in the considered case is less than for diagonal stripes.
Therefore the condition $n < 1/\lambda$ can be valid
for larger values of $n$,
resulting in additional bands inside the gap,
as it is seen from numerical results.

\subsection{Checkerboard structure}
As we have seen the spin-charge density structure
may be arranged  in vertical (horizontal) or diagonal
directions. Consider the solution with square symmetry.
 In the same approximation as before we find the solution of
system (\ref{ca}) in the form $m_{\pm}(\rr) = m (\rr_{\pm})$,
 $u_p(\rr) = {u}_p (\rr_{\pm})$, $v_{p} (\rr ) =
{v}_p (\rr_{\pm})$. Equations are decoupled and we
obtain
 \beqa
-iV_p\frac{d {u}_+}{dr_{\pm}} + m (r_{\pm}) {u}_- = E {u}_+\\
m^* (r_{\pm}) {u}_+  + i V_p \frac{d{u}_-}{d r_{\pm}} = E{u}_-,
\eeqa
with $V_p = 2t \sin p_x$, $r_{\pm}= (\pm x + y)/\sqrt{2}$.
The one "cross" solution has the form
\beq
m_+ = m_0 \tanh \frac{r_+}{\xi},\qquad
m_- =  m_0 \tanh\frac{r_-}{\xi},
\eeq
The spectrum $E$ and wave functions are found as above for the case
of stripes.
In the case of high doping the one kink solution
is transformed to the periodic structure
\[
 \langle S_z (\rr) \rangle \propto
(-1)^{x+ y} m_0 \cos[ \pi (\sqrt{\rho} -1 )x] \cos [ \pi (\sqrt{\rho} - 1)y],
\]
in which we considered the squared Fermi surface approximation with electron density
$\rho = |\Q|^2/ 2\pi^2$.

\section{Superconductivity and spin-charge modulation}
\subsection{Vortex solution}
 Consider pure superconducting state ($\Delta(\rr)\equiv 0$).
 The BdG equations are decoupled.
The first pair is
\beq
-i \V_{\p} \nabla_{\rr} u_{\p} (\rr) + \Delta_{\p}v_{\p} = \eps u_{\p}
\label{v1}
\eeq
\beq
\Delta_{\p}^* u_{\p} + i\V_{\p} \nabla_{\rr}v_{\p} (\rr) =\eps v_{\p}.
\label{v2}
\eeq

When  the filling $\rho$ is  close to 1, the Fermi surface
has nearly square form, therefore
$\V_{\p} \nabla_{\rr} \approx V_{\p} \partial/\partial r_{\pm}$, depending on signs $p_x,  \, p_y$
In this case the system of equations (\ref{v1}), (\ref{v2})
has the following vortex solution:
\beq
\Delta_{\p}(\rr ) = \Delta_p\frac{\sinh \frac{r_+}{\xi_s} + i \sinh \frac{r_-}{\xi_s}}
{\sqrt{\sinh^2 \frac{r_+}{\xi_s} + \sinh^2 \frac{r_-}{\xi_s} + 1}},
\eeq
where $\Delta_p = \Delta_0 (\cos(p_x) - \cos(p_y)) $.
For $r_- = 0$ the  order parameter has  a kink form $\Delta_p(\rr) \propto \tanh r_+ /\xi_s$.
For the case $r_+ = 0$ the order parameter acquires the phase:
$\Delta \propto  \exp( i \pi / 2)\tanh r_- /\xi_s$.
In the  diagonal direction $r_+ = r_-$ the solution
$\Delta_p(\rr) \propto \tanh r_+ /\xi /\sqrt{\tanh^2 r_+ /\xi_s  + 1} \exp(i \pi /4) $ has the
  phase $\pi /4$.
It is known that  in one-dimensional case  finite-band solutions of equations (\ref{v1}) - (\ref{v2})
are related to the soliton (kink) solutions of the  nonlinear Schrodinger equation (NSE).
Note, that
along the curve $ \sinh r_- /\xi_s = \alpha \cosh r_+ /\xi_s$  the order parameter
acquires the form of a general kink solution of   the NES:
$\Delta_p(\rr) \sim (i \alpha + \tanh x/\xi_s )/\sqrt{\alpha^2 + 1}$ with  the localized state in the gap
with the energy $E_0 = \Delta_p \alpha /\sqrt{\alpha^2 + 1}$.

\subsection{Coexistence of spin-charge structure and superconductivity}
Consider  solutions of  equations (\ref{ca}) in the superconducting region.

By  analogy with 1D case \cite{ma} we use ansatz:
\[
v_{\pm} = \gamma_{\pm} u_{\mp}
\]
which takes place in the uniform case.
The term $U\rho(\rr)/2$ in equations  can be eliminated by the shift
of wave functions $u,v \to u,v \exp i\Phi$, ${\bf V_p} \nabla \Phi
=U \rho(\rr )/2$. Considering $\eps(\p ) - \mu = 0$ on the Fermi surface
we obtain for the case
\[
m (\rr ) = \vert m(\rr) \vert e^{ i \varphi},\,
\Delta_{\p} (\rr) = \vert \Delta_{\p} (\rr)\vert e^{i\varphi_s},\,
\varphi,\, \varphi_s = const,
\]
the solution
$
\gamma_+ = \pm i e^{i(\varphi -\varphi_s )}, \,
\gamma_- = \pm i e^{-i(\varphi + \varphi_s)}$,
and the system (\ref{ca}) acquires the form

\beqa
-i\V_{\p}\nabla u_+ + \tilde{\Delta}(\rr)u_- = E u_+\label{se0}\\
\tilde{\Delta}^*(\rr)u_+ + i\V_{\p}\nabla u_- = Eu_-
\label{se}
\eeqa
with $\tilde{\Delta}(\rr) = (\vert m(\rr)\vert  \pm i\vert\Delta_{\p}\vert)
e^{i\varphi}$, and, as before, $m = m_{\pm}$, depending on the sign
of $p_y$. Equations (\ref{se0}), (\ref{se}) are exact provided that phases
$\varphi$,  $\varphi_s$ are constant or slowly varying in space functions.
We show that  inhomogeneity of the superconductor order parameter leads
to the origination of the antiferromagnetic order parameter.
Consider a 1D geometry  case: $ u = u(r_+)$, where assumption of constant phases is valid.
 The  solution of Eqs. (\ref{se0}), (\ref{se}) describing the coexistence of superconductivity
   and spin-charge  density ordering,  compatible  with self-consistent equations,
 has the form of two bound  solitons of the nonlinear Schrodinger equation, see, for example, \cite{fadtakh}
 \beq
 \tilde{\Delta}_{1,2} = \Delta_p \frac{\cosh (2 \kappa x + c_1) + \cosh (c_2 \pm 2 \beta i)/|\lambda| }
 {\cosh (2 \kappa x + c_1) + \cosh (c_2 )/|\lambda|},
 \eeq
 where $\pm \lambda ({\bf p})$ are positions of local levels inside the gap,
 $\kappa = \sqrt{\Delta_p^2 - \lambda^2}$, and $\exp i\beta = \lambda + i\kappa$.
 Eigenfunctions of equations (\ref{se0}), (\ref{se})  have the form \cite{bm}
 \beq
 u_{\pm}(x) \propto \sqrt{\tilde{\Delta}(x) (E^2 - \gamma^2(x))}\exp \left[\pm i\int^x
 \frac{\sqrt{(E^2 - \Delta_{p}^2)(E^2 - \lambda^2)}}{E^2 - \gamma^2(y)} dy \right],
 \label{wavefunction}
 \eeq
 where
 \beq
 \gamma (x) = \frac{1}{2} \frac{\partial}{\partial x} \ln \tilde{\Delta}(x).
 \eeq
For superconducting and spin order parameters we obtain
\beq
\Delta_{sc} = \Delta_p (1 - \Gamma \tanh a (\tanh(\frac{r_+}{\xi} + \frac{a}{2}) - \tanh (\frac{r_+}{\xi}- \frac{a}{2} ))),
\eeq
\beq
m = m_0 \Gamma \tanh a (\tanh(\frac{r_+}{\xi} + \frac{a}{2}) - \tanh (\frac{r_+}{\xi}- \frac{a}{2} ))
\eeq
where averaged over Fermi surface functions are defined as:  $\Gamma = <\Gamma_p>=<\Delta^2_p /(\Delta_p^2 + m_0^2)>_p $,  $\xi = <v_p/(\Delta_p \sqrt{\Gamma_p} \tanh a)>_p$, and we use the parametrization for the local level $\lambda$:
\beq
\lambda^2 = \frac{\Delta_p^2}{\Delta_p^2 + m_0^2}\left(m_0^2 + \frac{\Delta_p^2}{\cosh^2 a}\right)
\eeq
 Values of $m_0$,  $\Delta_{\p} = \Delta_d (\cos p_x - \cos p_y)$,
$\xi$ and a dimensionless parameter $a$ are defined by
the self-consistent conditions (\ref{s1}), (\ref{s2}).
The solution describes the spin-charge stripe in superconducting phase.

 \begin{figure}[tbph]
\begin{center}
\includegraphics[width=3.0in]{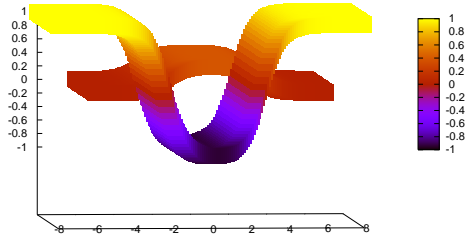}
\caption{ False color plot of the coexisting superconducting (downward) and antiferromagnetic stripe-like (upward) orders. The envelope functions are plotted in real space, $x$ and $y$ coordinates are measured  in units of correlation length $\xi$, $a=5$, $\Delta_d=1$, $m_0=0.2$.}
\end{center}
\end{figure}

The spin inhomogeneity generates the charge distribution $\delta \rho (\rr ) \propto
m^2(\rr )$. Note, that two-soliton solution in the similar form was used for describing
polaron-bipolaron states in the Peierls dielectrics \cite{brkir}.

The superconducting  correlation length  is increased in comparison to clean superconductor case $\xi_{sc}$  as $\xi = \xi_{sc} \sqrt{1+ \xi^2_{sc} /\xi^2_{AF}}$.

\section{Discussion}

We considered a simple self-consistent 2D model  on a squared lattice to describe different
states, including charge-spin structures, superconductivity, and their coexistence.
The origin of spin-charge periodic state (which is responsible for the pseudogap) is due to the existence of flat parallel
segments of the Fermi surface (nesting) at low hole doping concentrations.
Effects of commensurability
lead to a pinning of stripe structure at rational filling
points $|\rho - 1| = m/n$.
As a result,
there is  an exponentially small (for large $n$) decrease in the total energy of the order
$\delta E \sim \exp( - c\, n)$   at any commensurate point, stabilizing stripes, as in 1D systems.
For this reason, we think, stripes  are mostly observable  near $n = 8$ point ($|\rho- 1|= 1/8$).
An increase of doping leads to the decrease of flat segments of the Fermi surface and attenuation of spin-charge structure.

We found the solution describing the coexistence of superconductivity and stripes (28), (29).
The decrease (or a deviation from the homogenous value ) of the superconducting order parameter
generates the spin-charge periodic structure in this region.  Note, that due to symmetry
of Eqs. (25), (26)  (duality $\Delta \leftrightarrow i m$)
we can write the same equation, describing the origin of superconducting  correlations in the region
of a inhomogeneity of spin-charge density.  The situation is qualitatively similar to the 1D case \cite{ma}.
Experimental data in underdoped high-T$_c$ cuprates LSCO \cite{aeppli1} indicates that antiferromagnetic stripe-like
spin-density order can be induced by magnetic field perpendicular to the CuO planes in the interval of fields much smaller than
upper critical field H$_{c2}$ . The size of the magnetically ordered domains exceeds superconducting vortex's core size $\xi_s$ and the
inter-vortex distance in the Abrikosov's lattice.    Our present theoretical results demonstrate that this is indeed possible in the simple Hubbard
t-U-V model that we consider. In particular, the dimensionless parameter $a$ in Eqs. (28), (29)  is an independent variational parameter and depends
on the magnetic and superconducting coupling strengths \cite{ma}, as well as on the magnitude of the external magnetic field. Hence, the size $\sim a\times \xi$ of the
antiferromagnetic domain (see Fig. 2, upward red plane bump) can exceed the superconducting (and magnetic) Ginzburg-Landau correlation length $\xi$
when $a(H)>>1$. Previously coexistence of superconducting order and slow antiferromagnetic fluctuations was studied merely on the basis of a phenomenological
Ginzburg-Landau free energy functional approach in \cite{sachdev1}.
We note also, that equations Eqs. (25)-(26) can be simply extended to include d-density waves (DDW).
\section{Acknowledgements}
This work is in part supported by RFFI grant 12-02-01018-a.


\end{document}